\documentstyle[prd,aps,preprint]{revtex}
\tighten
\begin{document}

\title{\bf ON THE NEWTONIAN COSMOLOGY EQUATIONS WITH PRESSURE}

\author{J.A.S. Lima$^{1,3}$\footnote[1]{limajas@dfte-lab.ufrn.br}, 
V. Zanchin$^{2,3}$\footnote[2]{zanchin@het.brown.edu} and 
R. Brandenberger$^{3}$\footnote[3]{rhb@het.brown.edu.}} 
  
\smallskip

\address{~\\$^1$Departamento de F\'{\i}sica Te\'orica e Experimental, \\
     Universidade Federal do Rio Grande do Norte, 
     59072 - 970, Natal, RN, Brazil.}

\address{~\\$^2$Departamento de F\'{\i}sica, Universidade Federal de Santa Maria, \\97119-900, Santa Maria, RS, Brazil.}

\address{~\\$^3$Physics Department, Brown University, Providence, RI 02912, USA.}

\maketitle

\vskip 1.5cm
\begin{abstract}

\noindent The basic equations describing a Newtonian universe with uniform pressure are reexamined. We argue that in the semi-classical formulation  adopted in the literature the continuity equation has a misleading pressure gradient term. When this term is removed, the resulting
equations give the same homogeneous background solutions with pressure and the same evolution equation for the density contrast as are obtained using the full relativistic approach.

\end{abstract}
 
\vfill

\setcounter{page}{0}
\thispagestyle{empty}

\vfill

\noindent BROWN-HET-1063 \hfill        NOVEMBER 1996.

\noindent hep-ph/yymmdd \hfill Typeset in REV\TeX

\vfill\eject

\baselineskip 24pt plus 2pt minus 2pt

\section{Introduction}

Seventeen years after Einstein's seminal paper on relativistic cosmology,
Milne and McCrea\cite{MM 34} obtained Newtonian analogs to the expanding 
dust-filled FRW models of the universe, thereby initiating the so-called
Newtonian cosmology. Apart from some conceptual ambiguities present in 
these pressureless models, there is no doubt that such analogs 
are relevant both from a pedagogical and a methodological point of view, since they describe, in a rather simple way, several features of the presently observed universe. A natural generalization of these models 
should include pressure effects, as an attempt to construct analogs to the full class of FRW models. Unfortunately, such 
an extension cannot be analytically implemented, at least when the approach is
based uniquely on 
Newtonian concepts (absolute time, Euclidean space and ordinary gravity). The basic 
difficulty arises because a uniform pressure, in the Newtonian framework, does not play any dynamical or gravitational role,
at the level of the continuity, Euler and Poisson equations. A ``neo-Newtonian'' description, however, has been derived long ago by McCrea\cite{McCrea 51} himself, and in a more definite 
form (without any concepts from general relativity), by Harrison\cite{EH 65}. The later approach is based on the following set of equations:

``Continuity" equation:

\begin{equation}
{\partial \rho \over \partial t }+ \vec{\nabla_r}.(\rho + {p \over c^{2}})\vec{u} = 0\, ,
\label{eq:1}
\end{equation}

``Euler" equation:
\begin{equation}
{\partial \vec{u} \over \partial t} + \vec{u}.\vec{\nabla}_r\vec{u} = 
-\vec{\nabla}_r \psi - (\rho+{p \over c^{2}})^{-1}{\vec{\nabla}_r p}\, ,
\label{eq:2} \\
\end{equation}

``Poisson" equation:
\begin{equation}
 \nabla^2_r \psi= 4\pi G(\rho+3{p \over c^{2}}) \, ,  \label{eq:3}
\end{equation}
where $c$ is the velocity of light and the  quantities $\rho$, $p$, $\vec{u}$ and $\psi$ denote, respectively, the density, pressure, field velocity and  generalized gravitational potential of 
the cosmic fluid. As discussed in detail by Harrison\cite{EH 65}, all corrections of order $c^{-2}$ are based uniquely on arguments coming from special relativity. The price to pay is that the standard Eulerian equations 
are recovered  only in the limit $c \rightarrow \infty$. However, the 
main pedagogical aspect contained in the original equations for a dust-filled universe has 
been preserved, viz., all the fundamental Newtonian notions like absolute time, Euclidean space and gravity are also part of the enlarged theoretical context 
represented by the above modified equations.  

\section{A New ``Continuity" Equation}

As is widely known, the above set of equations admits a homogeneous and isotropic solution, i.e., $p=p(t)$ and $\rho=\rho(t)$. In this case, the fluid velocity is (a dot means partial time derivative)
\begin{equation}
\vec{u}= {\dot{a}\over a}\vec{r}\,  ,\label{eq:4}\\
\end{equation}
where the evolution of the scale factor $a(t)$ is governed by the Friedmann
equations, namely:
\begin{equation} 
{\ddot{a}\over a} = -{4\pi G \over3}(\rho +3{p \over c^{2}})\, ,\label{eq:5}\\
\end{equation}
\begin{equation}
 {\dot{a}^2\over a^2} =
 {8 \pi G\over3} \rho -{kc^{2}\over a^2} \, , \label{eq:6}
\end{equation}
with the continuity equation (\ref{eq:1}) reducing to the following form
\begin{equation}
{\partial \rho \over \partial t }+ 3{\dot{a}\over a}(\rho + {p \over c^{2}}) = 0\, .
\label{eq:7}
\end{equation}

   Equations (5)-(7) are exactly the same as obtained from Einstein's field equations, when one adopts comoving coordinates to describe a homogeneous and isotropic spacetime filled with a perfect simple fluid. Paradoxically, if one studies perturbation theory in the framework of equations (\ref{eq:1})--(\ref{eq:3}),
the evolution equation for the density contrast does not agree with the relativistic result in synchronous gauge. The expected result is obtained only for a
dust-filled universe ($p=0$). This  problem was first mentioned in a short note written in the appendix of the paper by Sachs and Wolfe\cite{SW 67}.
  
    The root of the problem seems to be 
closely related to equation 
(\ref{eq:1}). As we shall see, this equation corresponds to the correct conservation equation only in the homogeneous case. If inhomogeneities are taken into account, as for instance in the theory of small fluctuations, 
the pressure gradient appearing in Eq.(1) generates extra terms, leading to results 
which disagree with those resulting from the relativistic theory. To clarify this point, let us consider
the energy conservation equation in the form $dE + pdV =0$.  
The energy contained in a volume V with radius $r \sim a(t)$ is $E \sim {\rho c^{2} a^3}$, and since the volume is changing
we get $$
{d\rho \over dt }+ (\rho + {p \over c^{2}}){1\over a^3}{d\over dt}a^3 = 0\, .
$$

Using Eq. (\ref{eq:4}) we have $\vec{\nabla}_r.\vec{u}= 3\dot{a}/a$, and then it follows that $$
{d\rho \over dt} +(\rho+{p \over c^{2}})\vec{\nabla}_r.\vec{u} =0 \, .  
$$

Finally, by expanding the total derivative, ${d\over dt} = 
{\partial \over \partial t} + \vec{u}.\vec{\nabla}_r$,  the above 
equation may be rewritten as 
\begin{equation}
{\partial \rho\over \partial t} + \vec{\nabla}_r.(\rho\vec{u}) + 
 {p \over c^{2}}\vec{\nabla}_r.\vec{u} =0\, . \label{eq:8}
\end{equation}

This is the correct continuity equation in the modified Newtonian approach to
cosmology when pressure effects are included. It looks like a balance equation for the matter density. The source term (the last term
in Eq. (\ref{eq:8})), is related  to the work $dW= pdV$ needed to expand the 
fluid from the volume $V$ to $V+dV$. In fact  $$
{1\over V}{dW\over dt} = p {4\pi a^2 da\over{4\over3}\pi a^3 dt} = 
3{\dot{a}\over a}p = p\vec{\nabla}_r.\vec{u} \, . $$
where we used Eq.(\ref{eq:4}) in the last 
equality. This term has to be added
to the usual equation of fluid dynamics  to 
account for the work related to
the local expansion of the fluid.
It is worth noticing that in the limit of uniform pressure, $p=p(t)$, 
 Eq. (\ref{eq:8}) 
can always be rewritten in a form resembling Eq. (\ref{eq:1}), the form usually adopted in the literature\cite{EH 95} (Note that McCrae\cite{McCrae 51} originally wrote down the equation in the form of \ref{eq:8}!). However, although 
formally reducing to the same conservation equation for a homogeneous distribution, the perturbed equations that follows from each case are quite different. As we shall see, unlike 
Eq. (\ref{eq:1}), the new ``continuity" equation  
(\ref{eq:8}) is completely consistent with relativistic perturbation theory. In particular, this solves the contradiction pointed out by Sachs and Wolfe.

\section{Perturbation theory}

To study the evolution of small fluctuations in an expanding Newtonian universe, let us now consider the 
standard perturbation ansatz (in this section we take $c=1$):
\begin{eqnarray}
\rho &=& \rho_o(t)[1+\delta(\vec{r},t)] \, , \label{eq:9}\\
p & =& p_o(t) + \delta p(\vec{r},t)\, , \label{eq:10}\\
\psi &= &\psi_o(\vec{r},t)+ \varphi(\vec{r},t)\, , \label{eq:11}\\
\vec{u} &=& \vec{u}_o + \vec{v}(\vec{r},t) \, , \label {eq:12}
\end{eqnarray}
where the quantities carrying the subscript ``$o$'' represent the 
(background) homogeneous solution to the unperturbed equations (\ref{eq:2}), (\ref{eq:3}) and (\ref{eq:8}). The quantities $\delta$, $\delta p$, $\varphi$ and $\vec{v}$ are to be considered small (perturbations) when compared to the their respective background quantities. It should be recalled that in terms of the comoving
coordinates $\vec{q}$ we have $\vec{r}(t)= a(t)\vec{q}$ and 
$\vec{u}_o= \dot{a}(t)\vec{q}$. 
 
  Inserting the above expressions into Eqs. (\ref{eq:2}), (\ref{eq:3}) and (\ref{eq:8}), and linearizing the resulting equations, we 
get to first order in the perturbations
\begin{eqnarray}
& & \rho_o \left[{\partial \delta \over \partial t} + \vec{u}_o.\vec{\nabla}_r\delta\right] -
3{\dot{a}\over a}p_o\delta + 3{\dot{a}\over a} \delta p
+ (\rho +p)\vec{\nabla}_r.\vec{v} =0\, ,\label {eq:13} \\
& &{\partial\vec{v}\over \partial t} + (\vec{u}_o.\vec{\nabla}_r)\vec{v}+ {\dot{a}\over a}\vec{v}= -\vec{\nabla}_r \psi
 - (\rho+p)^{-1}{\vec{\nabla}_r \delta p}\, ,\label{eq:14} \\
& & \nabla_r^2\varphi = 4\pi G\rho_o( \delta+3{\delta p\over \rho_o})\, .
\label{eq:15}
\end{eqnarray}

 We now change to comoving coordinates $\vec{q} = \vec{r}/a$. Following standard lines\cite{PE 93}, we transform the partial time derivative of an arbitrary function $f$ at fixed $\vec{r}$, 
$\frac{\partial f}{\partial t}$\, , to the partial time derivative of $f$ at a fixed $\vec{q}$, which we write as $\dot{f}\,$.
The relation between these two partial time derivatives is
\begin{equation}
\left({\partial f\over \partial t}\right) = \dot{f} 
       -{\dot{a}\over a}\vec{q}.\vec{\nabla}f  \, , \label{eq:16}
\end{equation}
where  $\vec{\nabla} =a\vec{\nabla}_r$ is the gradient with respect to
$\vec{q}$ at fixed time.
 
For the sake of simplicity, let us also assume that the medium satisfies the equation of state 
$p=\nu \rho \, ,$ 
where  $\nu$ is a constant. For adiabatic perturbations, this means that $\delta p= v_s^{2}\delta\rho=v_s^{2}\rho_o\delta$, where
$v_s^{2}=\nu$ is the sound velocity. With this choice and using Eq. 
(\ref{eq:16}), Eqs. (\ref{eq:13})--(\ref{eq:15})
reduce to
\begin{eqnarray}
& &\dot{\delta}+{1+\nu\over a}\rho_o\vec{\nabla}.\vec{v}=0\, ,\label{eq:17}\\
& &\dot{\vec{v}} + {\dot{a}\over a}\vec{v}= -{1\over a}\vec{\nabla}\varphi       -{v_s^2\over (1+\nu)q}{\vec{\nabla}\delta} \, , \label {eq:18}\\
& & \nabla^2\varphi = 4\pi G (1+3\nu)\rho_o a^{2}\delta\, .\label{eq:19}
\end{eqnarray}

Finally, by eliminating the peculiar velocity
from Eqs.(17) and (18), and using (19), is readily seen that
\begin{equation}
\ddot{\delta} +2{\dot{a}\over a}\dot{\delta} -4\pi G(1+\nu)(1+3\nu)\delta
={v_s^{2} \over a^2}\nabla^2\delta\, . \label{eq:20}
\end{equation}
This is the differential equation governing the evolution of the density contrast, when
we describe a mass distribution with uniform pressure using the modified 
Newtonian equations. It coincides with 
the general relativistic equation (in the
synchronous and comoving gauge), for any value of the parameter 
$\nu$.
In particular, for scales bigger than the Jeans' length, when the Laplacian 
term on the right hand side may be neglected, Eq. (\ref{eq:20}) give rise 
to the same kind of gravitational instability usually derived in the relativistic treatment (see, for instance, Eq.(10.118) in Peebles' book\cite{PE 93}).

As an illustration of these results, we consider the unperturbed  
solution, 
$a(t) \sim t^{\frac {2}{3(1 + \nu)}}$, which is obtained when the field 
velocity coincides with the 
escape velocity (flat model, in the relativistic terminology). In this case,
$\rho_o \sim a^{-3(1 + \nu)}$, and for scales 
bigger than the Jeans' length, the 
growing mode of (\ref{eq:20}) scales as 
$\delta_+ \sim a^{1+ 3\nu}$. It thus follows, 
from (\ref{eq:19}), that the correction 
to the Newtonian potential is time-independent, regardless of the value of $\nu$. As a consequence, there is no
contribution to the temperature anisotropy of the cosmic background 
radiation due to the integrated Sachs-Wolfe effect. In fact, this effect is given by

\begin{equation}
\frac {\delta T} {T}({\bf n})\left|_{\rm ISW}\right. = \int_{i}^{f} d\tau {\dot \varphi} (\tau, {\bf n}(\tau_o - \tau))\, ,
\label{eq:21}\\
\end{equation}
where the integral is taken along the photon path from the last scattering surface to the observer\cite{CL 95}. Since the 
perturbed Newtonian potential is time 
independent for any value of $\nu$, it thus follows
that $\frac {\delta T}{T}\left|_{\rm ISW}\right. = 0$. This generalizes the well known result for a dust-filled universe. 
Note that this result (time independent Newtonian gravitational potential) is in agreement with what we expect from a Newtonian analysis. 

For completeness, we also derive the evolution equation for the density contrast, using the standard ``continuity" equation (\ref{eq:1}). Inserting the
perturbation ansatz in  Eqs. (\ref{eq:1})--(\ref{eq:3}) and following the same steps as above, we arrive at the following equation 
\begin{equation}
\ddot{\delta} +2{\dot{a}\over a}\dot{\delta} -4\pi G(1+\nu)(1+3\nu)\delta
+{\dot{a}\over a} \nu \vec{q}.\vec{\nabla}\dot{\delta} 
+{\ddot{a}\over a}\nu \vec{q}.\vec{\nabla}\delta 
={v_s^{2} \over a^2}\nabla^2\delta\, . \label{eq:22}
\end{equation}

Comparing to Eq. (\ref{eq:20}), we see that (22) has two additional terms (the last two 
on the LHS), and they have no counterpart in the relativistic theory of cosmological perturbations. Naturally, their ultimate origin is traced back to the pressure gradient term 
present in the standard ``continuity" equation. As a matter 
of fact, we
cannot eliminate these terms by simply rewriting the perturbed equations in the comoving coordinate system (where the background velocity 
is zero), as recently claimed by 
some authors\cite{CL 95}. As readily seen from (22), such a trick works only 
for a dust-filled universe($\nu=0$), when both ``continuity" equations 
reduce to the ordinary mass conservation law.    

\section{Conclusions}
We have shown that the correct relativistic equations for the cosmological background and for the density contrast
can be obtained from Newtonian cosmology even in the presence of nonvanishing pressure. This is achieved by a careful consideration of what a modified continuity equation should look like. The equation for the density contrast
agrees with the corresponding relativistic equation written in the synchronous and comoving gauge. Moreover, as a byproduct we get a Newtonian potential which is well defined independent of the fluid equation of state. In principle, these results imply that the domain of applicability of Newtonian cosmology may be enlarged to analyse some problems of structure formation even in the radiation dominated phase.  
  
\bigskip

\centerline{\bf Acknowledgments}

This work is partially supported by the Conselho Nacional de 
Desenvolvimento Cient\'{i}fico e Tecnol\'{o}gico- 
CNPQ (Brazilian Research Agency), and by the US Department of Energy under contract DE-FG0291ER40688, Task A .

\end{document}